# A Primary Frequency Control Strategy for Variable-Speed Pumped-Storage Plant in Power Generation Based on Adaptive Model Predictive Control

Zhenghua Xu, Changhong Deng*, and Qiuling Yang

*Abstract*—Variable-speed pumped-storage (VSPS) has great potential in helping solve the frequency control problem caused by low inertia, owing to its remarkable flexibility beyond conventional fixed-speed one, to make better use of which, a primary frequency control strategy based on adaptive model predictive control (AMPC) is proposed in this paper for VSPS plant in power generation. Under AMPC, the mechanical and electrical power of VSPS units are further coordinated such that stronger response to frequency variation can be provided. To make AMPC applicable, a prediction model with appropriate simplification is adopted and several equivalent control constraints are established to reflect VSPS unit's operating limits that can not be expressed directly due to the simplification. Based on a test system modified from IEEE 24-bus system, a case study is conducted through hardware-in-the-loop simulation to verify the effectiveness and feasibility of AMPC, and the results show that when AMPC is applied to VSPS plant, the maximum frequency deviation of the system under disturbance is reduced by up to 49.07%, compared with that when fixed-speed units are adopted, and up to 21.30%, compared with that when virtual inertia control is applied.

## I. Introduction

THE increasing penetration of converter-based renewable energy generation in power system is replacing conventional synchronous-machine-based power generation and reducing the system inertia, which makes grid frequency prone to large deviation when disturbance occurs and poses a challenge to primary frequency control (PFC)[1], [2]. Among the potential solutions, leveraging energy storage system (ESS) with appropriate frequency control strategy is an effective and economical option[3]. Pumped-storage is the most efficient and practical largescale ESS[4] and occupies the vast majority of global energy storage capacity, accounting for more than 90%[5]. The variable-speed pumped-storage (VSPS) unit employing doubly-fed induction machine (DFIM) is a new type of pumped-storage unit with the advantages of large capacity (up to 400MW[6]), fast response (power step change of 0.1 ~ 0.2 pu can be realized within 0.2s[6]), and high efficiency (typical overall efficiency is in the range of 70–85%[4]), which has been the most preferred choice for high power ratings[4], [7].

It is well-known that grid frequency dynamic is strongly related to the active power balance of power system[8]. Thus the potential of a facility to help maintain grid frequency stability depends on the active power flexibility it can provide.

For active power control in power generation, in addition to the fast response mentioned above, VSPS unit has another advantage of an extra degree of freedom in control. Fig. 1 shows the control process of VSPS unit and conventional fixed-speed pumped-storage (FSPS) unit based on synchronous machine (SM), where all the feedback are omitted for the sake of intuition. For FSPS unit, the only control input is the guide vane opening $G$, and the regulation of active power $P_e$ requires a slow process dominated by the dynamics of the mechanical power $P_m$ and power angle $\theta$. However, VSPS unit's $P_e$ can be fast regulated by excitation voltage $\boldsymbol{u_r}$, owing to the fast dynamics of converters and DFIM's electromagnetism. Besides, VSPS unit's $P_m$ can be regulated by $G$ independently at the same time. Therefore, VSPS unit has a total of two control inputs involved in the active power control. Fast response capability and an extra degree of freedom in control can give VSPS unit rich flexibility and thus great potential in helping solve the frequency control problem caused by low inertia.

At present, the researches on VSPS's PFC strategy are few and focus only on utilizing the fast response capability of VSPS unit's $P_e$, where the strategies work through $P_e$'s fast tracking of active power command $P^*$ composed of the dispatched command $P^*_{set}$ and the additional command $\Delta P^*$ related to grid frequency variation[6], [9]–[13], as shown in Fig. 1 (b). The droop control in [9] makes $\Delta P^*$ proportional to grid frequency deviation to make VSPS unit provide damping during PFC dynamic and share load change in steady-state, and in [10] the

This work was supported in part by the National Key Research and Development Program of China (2017YFB0903705).

Zhenghua Xu is with the school of electrical engineering and automation, Wuhan University, Wuhan, China (e-mail: xuzhenghua@whu.edu.cn).

Changhong Deng* is with the school of electrical engineering and automation, Wuhan, University, Wuhan, China (e-mail: dengch-whu@163.com).

Qiuling Yang is with the school of electrical engineering and automation, Wuhan University, Wuhan, China (e-mail: qiuling_yang@yeah.net).

Digital Object Identifier

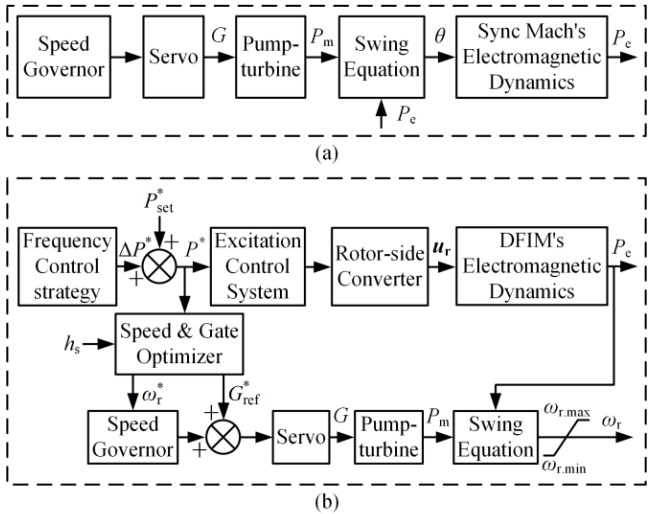

Fig. 1. Active power control process in generation mode
(a) FSPS unit; (b) VSPS unit

adaptive droop results in lower frequency deviations compared with fixed droop; the virtual inertia control (VIC) in [11] uses $\Delta P^*$ proportional to the derivative of grid frequency to make VSPS's response simulate the inertia response of SM; the robust adaptive control in [12] provides $\Delta P^*$ in terms of nonlinear control law and pulls grid frequency to reference model output adaptively; the control strategy in [13] generates $\Delta P^*$ by extracting a specific frequency band included in wind farm output through notch filter to compensate wind power fluctuation. Under these strategies, $P_m$ simply tracks $P_e$ according to swing equation under the action of speed governor and therefore the flexibility advantage brought by the extra degree of freedom in control is not fully utilized.

Although VSPS unit's $P_e$ and $P_m$ can be independently controlled by $u_r$ and $G$ respectively, they are related by swing equation and the balance between them affects the rotor speed $\omega_r$, as shown in Fig. 1 (b). Hence, for VSPS unit, there needs to be a coordination between $P_e$ that can be adjusted quickly and $P_m$ that responds slower, so as to keep $\omega_r$ within the allowable range $[\omega_{r.min}, \omega_{r.max}]$. Feedforward control proposed in [6] is widely used for the coordination, which is shown in Fig. 1 (b). When the feedforward control is applied, $P^*$ is sent to the excitation control system to regulate $P_e$ on the one hand, and to the optimizer of $\omega_r$ and $G$ on the other hand. The optimizer generates the rotor speed command $\omega_r^*$ and guide vane opening command $G_{ref}^*$, which can help achieve the optimal efficiency under $P^*$ and water head $h_s$. $G_{ref}^*$ acts on servo motor through feedforward control to achieve faster response of pump-turbine. However, this feedforward control only accelerates the tracking of $P_m$ to $P_e$ to a certain extent, and there is not enough coordination between $P_m$ and $P_e$, for both $P_e$ and $P_m$ always track the same command $P^*$, though they belong to different parts of VSPS unit with different dynamic characteristics. Considering both the electromagnetic and mechanical dynamics at the same time, nonlinear model predictive control (MPC) is adopted in [14] to regulate VSPS unit's power, which achieves good command tracking performance and robustness, but it does not consider VSPS unit's participation in PFC.

As for the application of MPC in energy storage system to participate in power system frequency control, there has been some research[15]–[17], where the main advantage of MPC is that it can comprehensively consider the overall dynamic characteristics and operating constraints of the controlled object, which gives MPC a promising application in coordinating the $P_e$ and $P_m$ of VSPS unit within the operating constraints.

In order to make better use of VSPS's flexibility during PFC, the following work is carried out in this paper.

Firstly, considering the further coordination between the electrical and mechanical power, the hydraulic coupling, and the nonlinearity of the plant dynamics, a PFC strategy for VSPS plant in power generation based on adaptive model predictive control (AMPC) is proposed, and the control scheme is elaborated.

Then, based on the analysis and simplification of electromagnetic dynamics, several equivalent control constraints are established for VSPS unit to reflect the operating limits that can not be expressed directly due to the abstraction of actual unit by the prediction model.

Next, a pattern-search-based design method of the control parameters in AMPC is introduced to help determine the horizon of prediction and control and the weights in the objective function.

Finally, a case study is conducted through hardware-in-the-loop (HIL) simulation to verify the effectiveness and feasibility of the proposed AMPC strategy, where IEEE 24-bus system is modified as the test system with a variety of power supplies and load and low inertia, and the results of pumped-storage plant employing FSPS units and VSPS units under AMPC and VIC are compared and discussed.

## II. DESIGN OF AMPC

In order to achieve the further coordination mentioned in I, there are two more things to consider. One is that there can be several VSPS units sharing a common penstock in a plant, which results in certain hydraulic coupling that affects each other's mechanical power regulation[18], [19]. The other is the stronger nonlinearity of VSPS unit's dynamics due to the greater variation of working conditions caused by the larger speed range.

With the good applicability to multi-input-multi-output system and the ability to consider the overall characteristics of the controlled object, MPC is well suited to solve the problems of further coordination and hydraulic coupling. However, due to the linear model required to ensure the efficiency of real-time computation, some measures must be taken to help MPC overcome the last problem of nonlinearity. Therefore, based on MPC and successive linearization, the adaptive model predictive control (AMPC) strategy is proposed for VSPS plant in power generation to participate in PFC, and the control scheme with variables in per unit is shown in Fig. 2, where the superscript "*" indicates the control instruction; the subscript "o" indicates the initial operating point; the subscript "$i$" is the serial number of a VSPS unit, and $i = 1, 2, …, n$, where n is the number of VSPS units in the plant; $\Delta$ represents the increment relative to the initial operating point; vector variables are shown in bold.

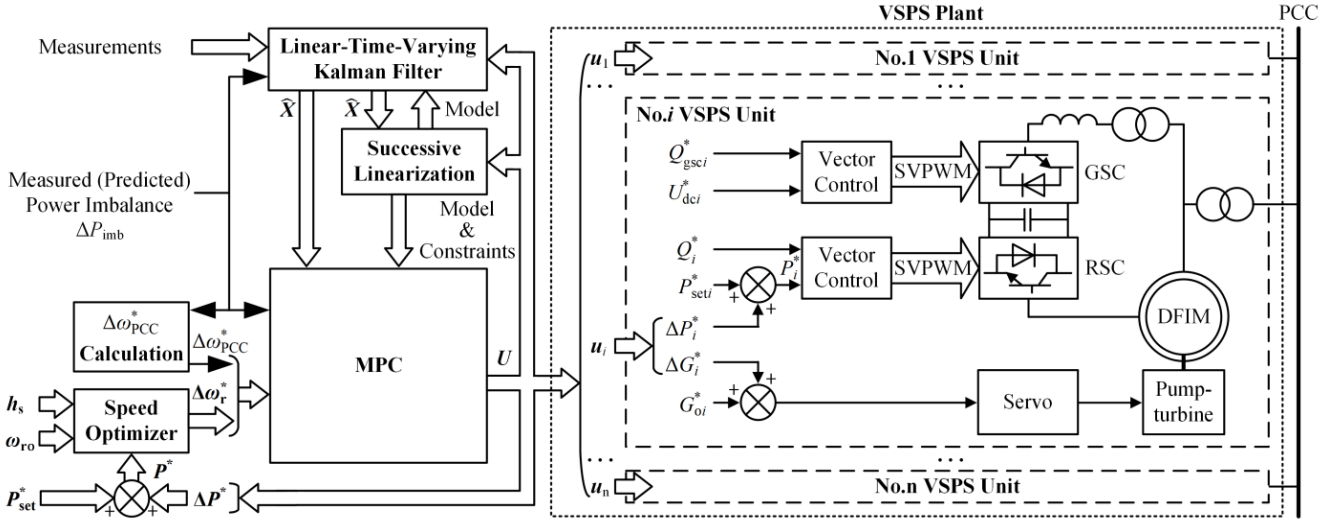

Fig. 2. The control sheme of AMPC strategy

For each VSPS unit in the plant in Fig. 2, vector control is used for power command tracking, which has become the mainstream by virtue of its good performance[7]. Under vector control, Grid-side converter is modulated to generate reactive power of $Q^*_{gsci}$ and keep the DC bus voltage at rated value of $U^*_{dci}$; Rotor-side converter is modulated to make the unit's active and reactive power track $P^*_i$ and $Q^*_i$ respectively. In this paper, it is assumed that VSPS plant does not participate in power system voltage control, and thus $Q^*_{gsci}$ and $Q^*_i$ are zero. $P^*_i$ is composed of dispatched command $P^*_{seti}$ and additional command $\Delta P^*_i$ generated by AMPC which is used to make VSPS unit respond to the angular frequency deviation at the point of common coupling (PCC), $\Delta\omega_{PCC}$. Besides, the speed governor is removed in each VSPS unit in Fig. 2 and $G_i$ is adjusted completely by AMPC via $\Delta G^*_i$ to achieve more flexible $P_m$ regulation.

The principle of AMPC computing and generating the control input $U$ is as follows: For each control cycle, firstly, Kalman filter[20] is used to obtain the estimation of current operating point, denoted as $\hat{X}$, based on the system measurements and control input feedback. Then, the parameters of prediction model and constraints are updated based on $\hat{X}$, so as to update the model and constraints in MPC and Kalman filter. Finally, based on the current model, constraints, $\hat{X}$ and command $\Delta\omega^*_{PCC}$ and $\Delta\omega^*_r$, MPC[21] is used to calculate $U$ series to achieve the optimal objective value within the prediction horizon, and issue the $U$ series within the control horizon. To achieve the reasonable distribution of power increment, $\Delta\omega^*_{PCC}$ is calculated from (1).

$$\Delta\omega^*_{PCC} = \frac{\sum_j S_j}{\sum_j (S_j/\delta_j) + F_{sys}} \Delta P_{imb} \quad (1)$$

where $S_j$ and $\delta_j$ are respectively the capacity and droop coefficient of $j$-th unit that participates in PFC; $F_{sys}$ is the system damping coefficient. $\Delta\omega^*_r$ can be obtained in a number of ways summarized in [7], among which the interpolation on the hill chart of pump turbine is adopted in this paper, as [22], [23] do, due to the low computation intensity and fast response.

The following describes the establishment of prediction model and objective function respectively.

*A. Prediction Model*

After the trade-off between simplicity and accuracy, the nonlinear prediction model with parameters in per unit is established first as Fig. 3 shows, where the electromechanical dynamic model established in [24] is adopted for VSPS unit to avoid significant time step reduction required by electromagnetic dynamics; the grid frequency dynamic is modelled by equivalent synchronous machine[8], for which the response from various components of the system are also modelled, including the prime mover and speed governor of synchronous unit[25], [19], the response dynamics and frequency control system of converter-based units[26], [27], and the self-regulation of load[28].

By linearizing the nonlinear prediction model shown in Fig. 3 at the current operating point and then discretizing it according to the prediction time step, the discrete prediction model for MPC can be obtained.

*B. Objective Function*

In order to accommodate the multiple control objectives of grid frequency stability and efficient operation of VSPS units and retain the advantage of quadratic programming, the objective function of step $k$ is established as shown in (2).

$$J_k = \frac{1}{2}\sum_{m=0}^{N-1}\left(p_{PCC}\left(\Delta\omega_{PCC}(k+m)-\Delta\omega^*_{PCC}(k)\right)^2 + \sum_{i=1}^{n}\left(p_i\left(\Delta\omega_{ri}(k+m)-\Delta\omega^*_{ri}(k)\right)^2\right)\right) \quad (2)$$

where $N$ is prediction horizon; $p_{PCC}$ and $p_i$ are the weight of $\Delta\omega_{PCC}$ error and $\Delta\omega_r$ of $i$th VSPS unit in the plant, and $i = 1,2,…,n$.

III. CONSTRAINTS FOR VSPS UNIT

Since the electromagnetic dynamics of VSPS unit are simplified to a first-order lag in the prediction model, the key variables such as stator and rotor current and rotor excitation voltage, as well as their limits, are unable to be expressed, for which equivalent constraints are established in this section so that AMPC can conform to the limits.

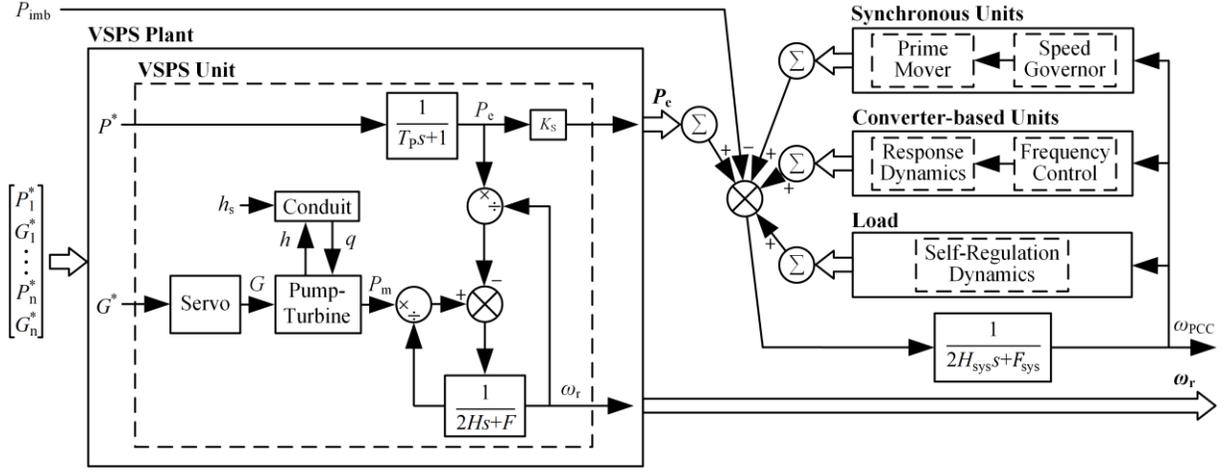

Fig. 3. Nonlinear prediction model

With the d-axis oriented to stator voltage and the stator flux variation and stator resistance ignored, VSPS unit's electromagnetic dynamics under motor convention after Park transformation can be transformed into (3)[29], where $u$ and $i$ are instantaneous voltage (V) and current (A); $R$ and $L$ are resistance (Ω) and inductance (H); the subscripts "d" and "q" denote the quantities of d-axis and q-axis; the subscripts "s", "r", "m" indicate the stator side, rotor side, and excitation quantities respectively; $U_s$ is the stator voltage amplitude; $\omega_{slp}$ is the slip angular frequency, and $\omega_{slp} = \omega_{PCC} - \omega_r$; $\sigma = 1 - L_m^2/(L_s L_r)$.

$$\begin{cases} u_{rd} = R_r i_{rd} + \sigma L_r \dfrac{di_{rd}}{dt} \underbrace{- \sigma L_r \omega_{slp} i_{rq} + \dfrac{L_m \omega_{slp} U_s}{L_s \omega_{PCC}}}_{C_{rd}} \\ u_{rq} = R_r i_{rq} + \sigma L_r \dfrac{di_{rq}}{dt} + \sigma L_r \omega_{slp} i_{rd} \\ i_{sd} = -(L_m/L_s) i_{rd} \\ i_{sq} = -(L_m/L_s) i_{rq} - U_s/L_s \omega_{PCC} \end{cases} \quad (3)$$

### A. Equivalent Constraint of Rotor Current Limit

According to (3), with the terms containing $R_r$ and $\sigma$ in rotor voltage expressions ignored due to the small rotor resistance and leakage reactance, the electrical power generated by VSPS unit can be expressed by rotor current, as shown in (4).

$$\begin{cases} P_e = -\dfrac{3}{2}(u_{sd} i_{sd} + u_{sq} i_{sq} + u_{rd} i_{rd} + u_{rq} i_{rq}) = \dfrac{3 L_m U_s \omega_r}{2 L_s \omega_{PCC}} i_{rd} \\ Q = -\dfrac{3}{2}(u_{sq} i_{sd} - u_{sd} i_{sq}) = -\dfrac{3 L_m U_s}{2 L_s} i_{rq} - \dfrac{3 U_s^2}{2 L_s \omega_{PCC}} \end{cases} \quad (4)$$

Substitute (4) into (5) that expresses rotor current limit, where $I_r$ is the rotor current amplitude and $I_{rN}$ is the rated value of $I_r$, and then linearize (5) to obtain the equivalent constraint of rotor current limit shown in (6), where $K_{ir} = L_s^2 \omega_{PCCo}^2 P_{eo}^2 / L_m^2 \omega_{ro}^2 U_s^2 I_{ro}$.

$$I_r = \sqrt{i_{rd}^2 + i_{rq}^2} \leq I_{rN} \quad (5)$$

$$\dfrac{3 K_{ir}}{2 P_{eo}} \Delta P_e + \left(-\dfrac{U_s^2}{I_{ro} \omega_{PCCo}^3 L_m^2} + \dfrac{K_{ir}}{\omega_{PCCo}}\right) \Delta \omega_{PCC} - \dfrac{K_{ir}}{\omega_{ro}} \Delta \omega_r \quad (6)$$
$$\leq I_{rN} - I_{ro}$$

### B. Equivalent Constraint of Stator Current Limit

Substitute (4) into the stator current expressions in (3) to eliminate rotor current, and then the stator current can be expressed by electrical power and rotor speed, as shown in (7).

$$i_{sd} = \dfrac{2 P_e}{3 U_s \omega_r}, i_{sq} = \dfrac{2Q}{3 U_s} = 0 \quad (7)$$

Substitute (7) into (8) that expresses stator current limit, where $I_s$ is the stator current amplitude and $I_{sN}$ is the rated value of $I_s$, and then linearize (8) to obtain the equivalent constraint of stator current limit shown in (9).

$$I_s = \sqrt{i_{ds}^2 + i_{qs}^2} = \dfrac{2 P_e}{3 U_s \omega_r} \leq I_{sN} \quad (8)$$

$$2 \Delta P_e - 3 I_{sN} U_s \Delta \omega_r \leq -2 P_{eo} + 3 I_{sN} U_s \omega_{ro} \quad (9)$$

### C. Equivalent Constraint of Rotor Voltage Limit

#### 1) Constraint for Steady-state

Due to the negligible rotor resistance and leakage inductance, the amplitude of rotor excitation voltage in steady-state can be characterized by the absolute value of coupling term in d-axis component $|C_{rd}|$. Fig. 4 depicts the characterization of $|C_{rd}|$ to $U_r$ within the rated operating range with the VSPS unit parameters in [30], and it shows that $|C_{rd}|$ can characterize 0.95 ~ 1 $U_r$. Therefore, the rotor voltage limit can be rewritten as (10), where $r_{Ur}$ is the correction factor and is related to specific unit parameters. Under the parameter of [30], $r_{Ur}$ is taken as 1/0.95.

$$U_r \approx r_{Ur} |D_{rd}| \leq U_{rN} \quad (10)$$

Linearize (10) and then the equivalent constraint of steady-state rotor voltage limit can be obtained as shown in (11).

$$(r_{Ur} L_r U_s - U_{rN} L_m) \Delta \omega_{PCC} - r_{Ur} L_r U_s \Delta \omega_r$$
$$\leq -(r_{Ur} L_r U_s - U_{rN} L_m) \omega_{PCCo} + r_{Ur} L_r U_s \omega_{ro} \quad (11)$$

#### 2) Constraint for dynamic

According to (3) and (4), the VSPS unit's $P_e$ is controlled only by $u_{rd}$, and the linear model from $u_{rd}$ to $P_e$ is shown in (12), where $K_{Ur2P} = U_s L_m \omega_{ro}/(R_r L_s \omega_{PCCo})$; $T_r = \sigma L_r / R_r$; $s$ is Laplace operator.

$$\dfrac{\Delta P_e(s)}{\Delta u_{rd}(s)} = \dfrac{K_{Ur2P}}{T_r s + 1} \quad (12)$$

For the step excitation of $\Delta P^*$, the relationship between the maximum value of $\Delta u_{rd}$ and the amplitude of $\Delta P^*$ is shown in (13).

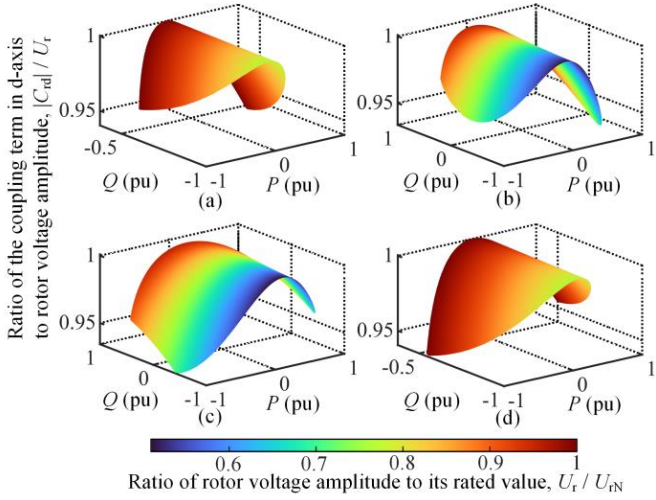

Fig. 4. Characterization of the d-axis compensation component for rotor voltage amplitude in rated operating range.
(a) $\omega_r = 0.85$; (b) $\omega_r = 0.9$; (c) $\omega_r = 1.1$; (d) $\omega_r = 1.15$.

$$\frac{\max(\Delta u_{rd})}{|\Delta P^*|} = K_{P2Ur} = \begin{cases} K_{Ur2P} \cdot \dfrac{T_r}{T_P}, & T_P < T_r \\ K_{Ur2P}, & T_P \geq T_r \end{cases} \quad (13)$$

In terms of the equivalent constraint analysis of rotor voltage limit in steady-state, the rotor voltage amplitude is mainly composed of the d-axis components, so $\Delta u_{rd}$ can be used to approximate the dynamic variation of rotor voltage amplitude and the equivalent constraint of rotor voltage limit for dynamic can be expressed as (14).

$$K_{P2Ur}\Delta P^* \leq U_{rN} - U_{ro} \quad (14)$$

## IV. DESIGN OF AMPC PARAMETERS

Parameters to be set in AMPC include the horizons of prediction and control and the weights in the objective function, which can be set in terms of the pattern-search-based algorithm shown in Fig. 5, where $K$ is the vector composed of all the control parameters mentioned above; $m_1$ and $m_2$ are respectively the expansion and contraction multiples of search grid, $m_1 > 1$, $0 < m_2 < 1$; $\varepsilon$ is the convergence threshold for mesh size; the detailed generation method of search grid can be found in [31]; the initial parameters take empirical values; $J_{\min}(K_i)$ is obtained by solving the ordinary MPC problem for the plant at initial operating point according to the prediction model, objective function and constraints established above.

## V. CASE STUDY

The test system is modified from IEEE 24-bus system[32] where the power grid, rated capacity and distribution of generation, peak value and distribution of load remain the same, but the structure of generation and load is modified as shown in Fig. 6. In the test system, there are two 250MW units of FSPS or VSPS in the plant at bus-13 with the mechanical parameters taken from [23] and the parameters of DFIM for VSPS taken from [30]; the wind and solar plants work in maximum power point tracking mode without participating in PFC; the battery unit adopts droop control to participate in PFC; the inertia and damping of the system are $H_{sys0}=54.39$ (s), $D_{sys0}=0.2$ (pu) respectively (The base value for capacity is 100MVA, and the pumped-storage units at bus-13 are not included). For comparison with AMPC, VIC is selected as the control, where the parameters are optimized based on the method in [33] to minimize the maximum grid frequency deviation as much as possible within the constraints.

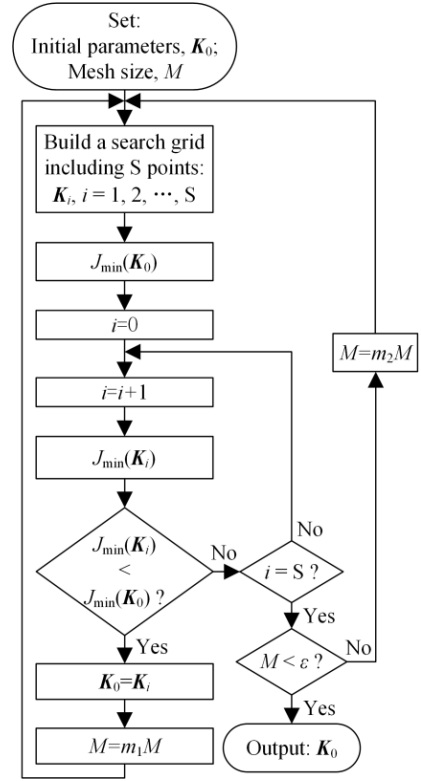

Fig. 5. Design algorithm of AMPC parameters

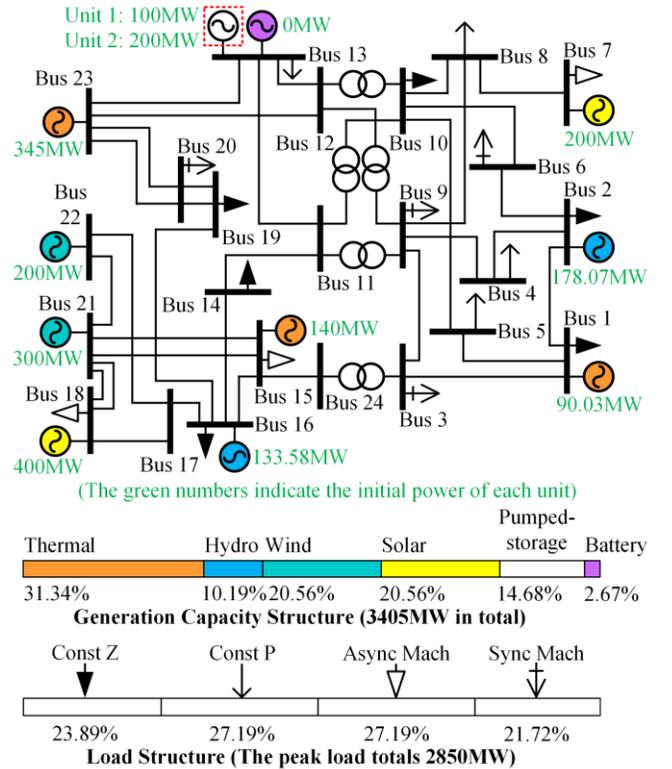

Fig. 6. Test sytem.

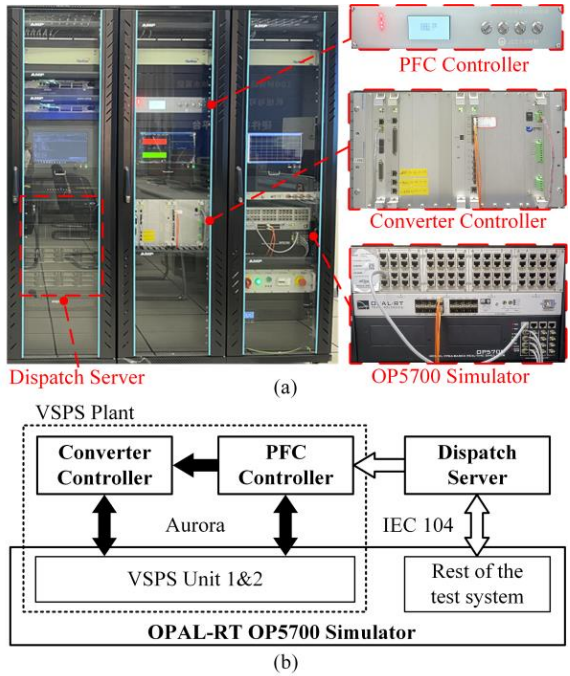

Fig. 7. HIL simulation platform.
(a) Physical hardware; (b) Architecture.

The simulation is conducted on the HIL simulation platform shown in Fig. 7, where the primary system model of the test system with detailed electromagnetic and mechanical dynamics is simulated in real-time with the time step of 50μs in the OP5700 simulator; the converter controller is responsible for the vector control of the converters of VSPS units in the plant, and interacts with the simulator and PFC controller via high-speed fiber-optic communication based on Aurora protocol; the PFC controller runs the frequency control strategy written on its DSP of TMS320F240; the dispatch server works as the dispatch center and delivers dispatching commands based on IEC 104 protocol; the measurements within the VSPS plant are sent directly to the converter controller and PFC controller and that out of the plant is first collected by dispatch server and then forwarded to the controllers.

At time zero, the load of bus line 5, 8 and 14 is increased by 100MW, totaling 300MW, and the simulation results of grid frequency variation and the response of pumped-storage are shown in Fig. 8. According to Fig. 8 (a), VSPS under VIC or AMPC can both significantly raise the nadir of $\omega_{PCC}$, compared with FSPS, and the nadir of $\omega_{PCC}$ is increased by 35.28% and 49.07% respectively. The improvement brought by VSPS comes from more flexible response of electrical power $P_e$. As shown in Fig. 8 (c) and (d), before $\omega_{PCC}$ reaches the nadir, VSPS can provide a larger $P_e$ response more quickly under both strategies to contribute more electric energy and support the stability of grid frequency. The flexible electrical power response of VSPS comes from the conversion of rotor rotating kinetic energy and the support of pump turbine mechanical power. For the former, the maximum variation of VSPS and FSPS units' rotor speed $\omega_r$ before $\omega_{PCC}$ nadir is about 0.05pu as shown in Fig. 8 (b), and i.e., the rotational kinetic energy that they convert into electrical energy is roughly equal. Although the total speed range of VSPS is usually up to ±10~20%, the $\omega_r$ achieving the optimal efficiency generally decreases with the decrease of load level, and the minimum speed only keeps a margin of about 5% based on the converter capacity[34]. Therefore, VSPS's advantage of larger speed range does not help much in this case, and it is the fast response of mechanical power $P_m$ that provides the main support for VSPS's $P_e$ response, as shown in Fig. 8 (e) and (f), where VSPS provides a $P_m$ response about twice as large as FSPS's before $\omega_{PCC}$ nadir.

Compared with VIC, AMPC can further raise the nadir of $\omega_{PCC}$ by 21.3%, as shown in Fig. 8 (a), and according to Fig. 9, each VSPS unit has at least one operating constraint close to being violated, which proves the accuracy of equivalent constraints and indicates that AMPC and VIC have both utilized the flexibility of VSPS to the best of their ability. With the importance of $P_m$ response having been highlighted above, it's not hard to find out that AMPC's better performance lies in extracting more flexibility from VSPS through greater coordination of $P_m$ and $P_e$: After load disturbance, as shown in Fig. 9 (d), AMPC adjusts VSPS's guide vane opening $G$ to the maximum at the maximum rate, which results in the strongest response of $P_m$ at the first time, as shown in Fig. 8 (e) and (f), where the $P_m$ response of VSPS under AMPC reaches a maximal value rapidly after disturbance while the $P_m$ under VIC is still climbing slowly. The stronger response of $P_m$ makes VSPS able to provide stronger $P_e$ response that acts directly on grid frequency dynamics, and VSPS's $P_e$ increment under AMPC is almost twice that under VIC when $P_e$ reaches the first maximal value and about 1.3 times at the second maximal value, as shown in Fig. 8 (c) and (d). The stronger $P_e$ response effectively prevents $\omega_{PCC}$ from falling and finally a smaller maximum deviation of $\omega_{PCC}$ is achieved. On the other hand, as shown in Fig. 9 (d), under VIC, $G$ changes more slowly under the regulation of speed governor and feedforward control for $\omega_r$ changes little at the beginning due to the slow mechanical process of rotor dynamics and the $G^*_{ref}$ issued by feedforward control is calculated based on steady-state efficiency rather than dynamic performance. Thus $P_m$ responds more slowly under VIC, as shown in Fig. 8 (e) and (f), though its maximum is about the same as that under AMPC. Limited by the speed range and relatively slow $P_m$ response, VSPS's $P_e$ responds slower and weaker under VIC, which results in larger maximum deviation of $\omega_{PCC}$.

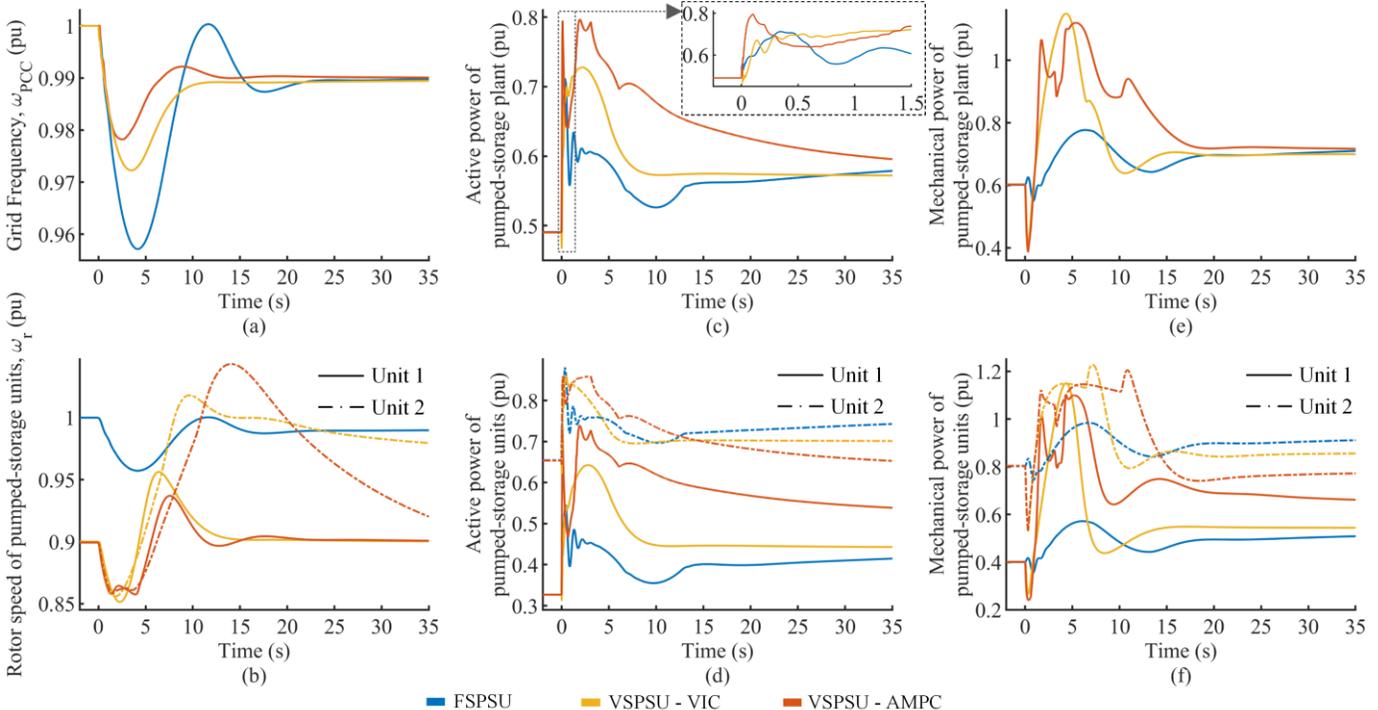

Fig. 8. Grid frequency variation and the reponse of pumped-storage

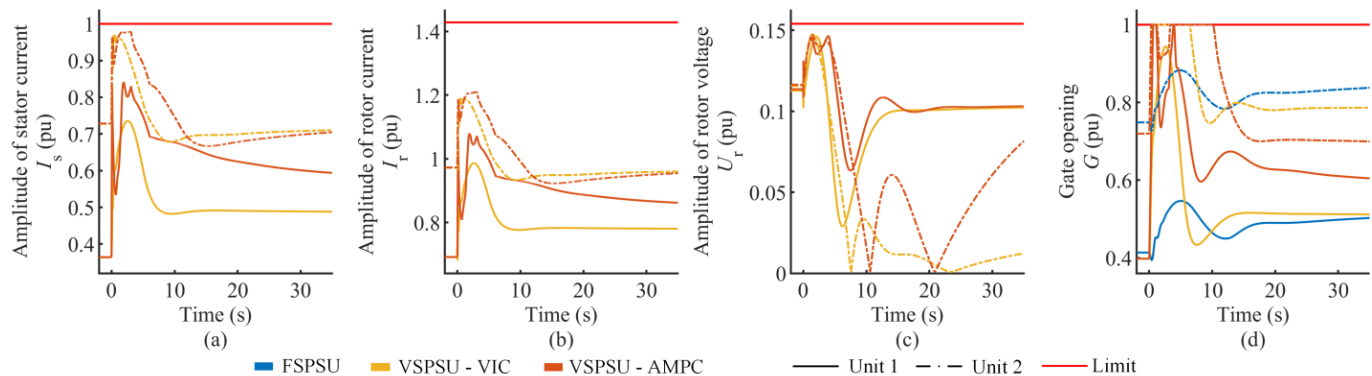

Fig. 9. Curves of the variables corresponding to operating constraints

## VI. Conclusion

The proposed AMPC strategy for VSPS plant in power generation takes into account the overall dynamic characteristics of the plant to overcome hydraulic coupling and achieve further coordination between electrical and mechanical power so that more flexibility of VSPS units can be leveraged to provide stronger response to grid frequency variation during PFC and thus grid frequency stability is enhanced. Besides, the adaptive capability of AMPC help get over the stronger nonlinearity of VSPS unit's dynamics due to the greater variation of working conditions caused by larger speed range, which is another factor for the good performance of AMPC.

The equivalent constraints established in this paper can accurately reflect the operating limits of VSPS units' electromagnetic dynamics to ensure safe operation, and in this way, the complexity of prediction model can be avoided, which enhances the feasibility of AMPC, for there can be multiple VSPS units to be controlled based on real-time prediction and optimization.

Good performance as AMPC achieves in this paper, there are still room for improvement, since the prediction and optimization of AMPC are only based on the current operating point without considering subsequent changes, and the robustness of AMPC remains to be studied.

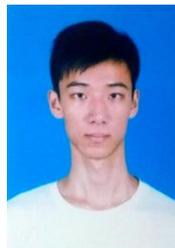

**Zhenghua Xu** received the B.Sc. degree and is currently working toward the M.S. degree in electrical engineering at the School of Electrical and Automation, Wuhan University, Wuhan, China.

His research interest areas include power system frequency control, modeling and control of doubly-fed variable speed pumped storage unit.

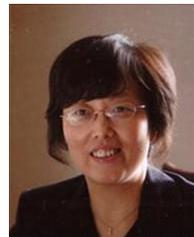

**Changhong Deng** is currently a professor of the School of Electrical Engineering and Automation, Wuhan University, Wuhan, China.

Her interest areas include security and stability analysis and control of large power grid, integration and control of renewable energy, intelligent control of power grid, optimization control of power grid, planning, modeling and digital simulation of large power system.

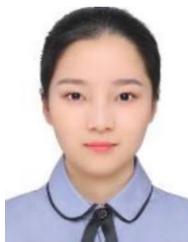

**Qiuling Yang** was born in Chongqing Province, China, in 2000. She received the B.S. degree inelectrical engineering from Sichuan University, Chengdu, China, in June 2020. She is currently working toward the M.S. degree in electrical engineering at the School of Electrical and Automation, Wuhan University, Wuhan, China.

Her research interests include new energy access and smart grid.